\documentclass[nofootinbib,prd,twocolumn,showpacs,showkeys,preprintnumbers]{revtex4-1}
\usepackage{hyperref,amssymb,amsmath,mathrsfs,bm,graphicx}
\begin{document}
\title {Stability of the isotropic pressure condition.}
\author{L. Herrera}
\email{lherrera@usal.es}
\affiliation{Instituto Universitario de F\'isica
Fundamental y Matem\'aticas, Universidad de Salamanca, Salamanca 37007, Spain}
\date{\today}
\begin{abstract}
 We investigate the conditions for the (in)stability of the isotropic pressure condition  in collapsing   spherically symmetric,  dissipative fluid distributions.   It is found that dissipative fluxes,  and/or energy density inhomogeneities and/or the appearance of shear in the fluid flow, force any initially isotropic configuration  to abandon such a condition, generating   anisotropy in the pressure. To reinforce this conclusion we also present some arguments  concerning the axially symmetric case. The consequences ensuing our results are analyzed.
 \end{abstract}
\date{\today}
\maketitle

\section{Introduction}
In theoretical physics it is usual to resort to different kinds of assumptions  in order  to solve (almost) any specific problem.  Assumptions are restrictions imposed to simplify the problem under consideration,  reflecting  some of the  essential aspects of the systems. Since all physical systems are subject to fluctuations,  those  essential aspects are as well. Accordingly  the following questions naturally arise in the study of almost any physical problem, namely:
\begin{itemize}
\item Is any result obtained under the assumption $A$ similar to that obtained under the ``quasi--assumption'' $A+\epsilon$ (where $\epsilon<<1$)? This question concerns the  stability of the result.
\item  Under which conditions does assumption $A$  remain valid all along the evolution of the system? This question concerns the stability of the assumption itself.
\end{itemize}

In this paper we endeavour to answer  the questions above, in relation to the isotropic pressure condition.

For many years, both in the Newtonian and the relativistic regime, the isotropy of the pressure (the Pascal principle) has been a common (and a fundamental) assumption in the study of stellar structure and evolution. Therefore the two questions above deserve to be answered for the isotropic pressure condition. 

The first question, concerning the stability of the result,  has a known answer. Indeed,    let us  recall that even a small pressure anisotropy may lead to results drastically different from the ones obtained by assuming isotropic pressure, due to the possible appearance of crackings  in the fluid distributions produced by the presence  of arbitrarily small pressure anisotropy  \cite{crak}. Thus, the stability of a specific result against small deviations from the isotropic pressure condition is not assured in general, and should be checked in each case.

Here, we shall focus  on the question concerning the stability of the isotropic pressure condition, i.e. under which conditions such an assumption remains valid all along the evolution?
More specifically, we endeavor   to  answer  the following  (related) questions:
\begin{itemize}
\item What   physical properties of the fluid distribution are related  (and how) to the appearance of pressure anisotropy in an initially isotropic fluid?
\item Under which conditions does an initially isotropic  configuration remain isotropic all along its evolution (stability problem)?
\end{itemize}

The relevance of the problem under consideration is illustrated, on the one hand, by the fact that many important results concerning relativistic fluids rely on the Pascal principle, and on the other hand,  by the fact that pressure anisotropy is expected to be produced by  physical processes usually present in very compact objects. This in turn explains  the renewed interest in self-gravitating systems with anisotropic pressure  observed  in recent years. Indeed, the number of papers devoted to  this issue  is so large that we ask for the indulgence of the reader for not being exhaustive with the corresponding bibliography.  Just as a small partial sample, let us mention the review paper \cite{rev} with a comprehensive  bibliography until 1997, and   some of the  recent works that have appeared so far in the current ($2020$)  year \cite{an1, an2, an3, an4, an5, an6, an7, an8, an9, an10, an11, an12, an13, an14, an15, an16,  an17, an19, an20, an21, an22, an23, an25, an26, an27, an28, an28bis, an29, an30, an31, an32, an33, an34, an35, an36, an37}.

Our approach  heavily relies on a  differential  equation  relating the Weyl tensor to different physical variables. It is an evolution equation containing time derivatives of  those variables. This equation was first  derived in \cite{ellis1, ellis2, ellis3} for configurations without any specific symmetry;  afterwards it was  reobtained and used in different contexts (see for example \cite{12, she, inh}).

We consider general fluid distributions endowed with anisotropic pressure and dissipating energy during its evolution. The specific physical (microscopic) phenomena behind these fluid characteristics are not discussed here; instead we  are concerned only by the macroscopic (hydrodynamic) manifestations of those phenomena. 

As we  see, only a highly unlikely  cancellation of terms containing the heat flux, the energy--density inhomogeneity and the shear of the fluid, could ensure that the pressure isotropy condition remains valid all along the evolution. To complement our discussion, we also  consider the axially symmetric case.

The manuscript is organized as follows: In the next section we introduce all the variables and  conventions used throughout the paper, for the spherically symmetric case. In sections III we   briefly present the  basic differential equation our study is based upon.  Then with all these elements we tackle in section IV the problem of identifying the conditions required for the pressure isotropy assumption  to remain valid all along the evolution of the system, and to identify the physical causes of the departure from such a condition. In order to strengthen further our case, we  expose some arguments concerning  the axially symmetric case  in section V. A  summary of the obtained results and a discussion on their potential consequences,  are presented in section VI. Finally an appendix with the expressions of Einstein equations and conservation equations for the spherically symmetric case  is included.

\section{ENERGY--MOMENTUM TENSOR,  RELEVANT VARIABLES AND FIELD EQUATIONS}
Let us consider a spherically symmetric distribution  of collapsing
fluid, non--necessarily  bounded. The fluid is
assumed to be locally anisotropic (principal stresses unequal) and undergoing dissipation in the
form of heat flow (to model dissipation in the diffusion approximation). 

Choosing comoving coordinates, the general metric can be written as
\begin{equation}
ds^2=-A^2dt^2+B^2dr^2+R^2(d\theta^2+\sin^2\theta d\phi^2),
\label{1m}
\end{equation}
where $A$, $B$ and $R$ are functions of $t$ and $r$ and are assumed
positive. We number the coordinates $x^0=t$, $x^1=r$, $x^2=\theta$
and $x^3=\phi$.

The matter energy-momentum tensor $T_{\alpha\beta}$ 
has the form
\begin{eqnarray}
T_{\alpha\beta}&=&(\mu +
P_{\perp})V_{\alpha}V_{\beta}+P_{\perp}g_{\alpha\beta}+(P_r-P_{\perp})\chi_{
\alpha}\chi_{\beta}+q_{\alpha}V_{\beta}\nonumber \\&+&V_{\alpha}q_{\beta}, \label{3}
\end{eqnarray}
where $\mu$ is the energy density, $P_r$ the radial pressure,
$P_{\perp}$ the tangential pressure, $q^{\alpha}$ the heat flux describing dissipation in the diffusion approximation, $V^{\alpha}$ the four velocity of the fluid,
and $\chi^{\alpha}$ a unit four vector along the radial direction. These quantities
satisfy
\begin{equation}
V^{\alpha}V_{\alpha}=-1, \;\; V^{\alpha}q_{\alpha}=0, \;\; \chi^{\alpha}\chi_{\alpha}=1,\;\; \chi^{\alpha}V_{\alpha}=0. 
\end{equation}

We do not explicitly add dissipation   in the free streaming approximation, bulk viscosity and/or shear viscosity to the system because they
can be absorbed into the energy density $\mu$, and the radial and tangential pressures, $P_r$ and
$P_{\perp}$, of the collapsing fluid.

Alternatively, we may write the energy--momentum tensor in its canonical form:
\begin{equation}
T_{\alpha \beta} = {\mu} V_\alpha V_\beta + P h_{\alpha \beta} + \Pi_{\alpha \beta} +
q \left(V_\alpha \chi_\beta + \chi_\alpha V_\beta\right) \label{Tab}
\end{equation}
with
$$ P=\frac{P_{r}+2P_{\bot}}{3}, \qquad h_{\alpha \beta}=g_{\alpha \beta}+V_\alpha V_\beta,$$

$$\Pi_{\alpha \beta}=\Pi\left(\chi_\alpha \chi_\beta - \frac{1}{3} h_{\alpha \beta}\right), \qquad \Pi=P_{r}-P_{\bot}.$$

Since we assume that our observer is comoving with the fluid then
\begin{eqnarray}
V^{\alpha}=A^{-1}\delta_0^{\alpha}, \;\;
q^{\alpha}=qB^{-1}\delta^{\alpha}_1, \;\;
\chi^{\alpha}=B^{-1}\delta^{\alpha}_1, \label{5}
\end{eqnarray}
where $q$ is a function of $t$ and $r$.

The four--acceleration $a_{\alpha}$ and the expansion $\Theta$ of the fluid are
given by
\begin{equation}
a_{\alpha}=V_{\alpha ;\beta}V^{\beta}, \;\;
\Theta={V^{\alpha}}_{;\alpha}, \label{4b}
\end{equation}
and its  shear $\sigma_{\alpha\beta}$ by
\begin{equation}
\sigma_{\alpha\beta}=V_{(\alpha
;\beta)}+a_{(\alpha}V_{\beta)}-\frac{1}{3}\Theta h_{\alpha \beta}.\label{4a}
\end{equation}

From  (\ref{4b}) with (\ref{5}) we have for the  four--acceleration and its scalar $a$,
\begin{equation}
a_1=\frac{A^{\prime}}{A}, \;\; a^2=a^{\alpha}a_{\alpha}=\left(\frac{A^{\prime}}{AB}\right)^2, \label{5c}
\end{equation}
where $a^\alpha= a \chi^\alpha$,
and for the expansion
\begin{equation}
\Theta=\frac{1}{A}\left(\frac{\dot{B}}{B}+2\frac{\dot{R}}{R}\right),
\label{5c1}
\end{equation}
where the  prime stands for differentiation with respect to $r$
and the dot stands for differentiation with respect to $t$.

With (\ref{5}) we obtain
for the shear (\ref{4a}) its non zero components
\begin{equation}
\sigma_{11}=\frac{2}{3}B^2\sigma, \;\;
\sigma_{22}=\frac{\sigma_{33}}{\sin^2\theta}=-\frac{1}{3}R^2\sigma,
 \label{5a}
\end{equation}
and its scalar
\begin{equation}
\sigma^{\alpha\beta}\sigma_{\alpha\beta}=\frac{2}{3}\sigma^2,
\label{5b}
\end{equation}
where
\begin{equation}
\sigma=\frac{1}{A}\left(\frac{\dot{B}}{B}-\frac{\dot{R}}{R}\right).\label{5b1}
\end{equation}
Then, the shear tensor can be written as
\begin{equation}
\sigma_{\alpha \beta}= \sigma \left(\chi_\alpha \chi_\beta - \frac{1}{3} h_{\alpha \beta}\right).
\label{sh}
\end{equation}

We can define the velocity $U$ of the collapsing
fluid  as the variation of the areal radius $R$ as
measured from its area, with respect to proper time, i.e.
\begin{equation}
U=\frac{\dot R}{A}. \label{19}
\end{equation}

\subsection{ Weyl tensor}

In general the Weyl tensor $C^{\rho}_{\alpha \beta \mu}$ may be   defined through  its electric and magnetic parts. However in the spherically symmetric case the magnetic  part vanishes identically, and the electric  part of  the  Weyl tensor is defined by
\begin{equation}
E_{\alpha \beta} = C_{\alpha \mu \beta \nu} V^\mu V^\nu,
\label{elec}
\end{equation}
with the following nonvanishing components
\begin{eqnarray}
E_{11}&=&\frac{2}{3}B^2 {\cal E},\nonumber \\
E_{22}&=&-\frac{1}{3} R^2 {\cal E}, \nonumber \\
E_{33}&=& E_{22} \sin^2{\theta}, \label{ecomp}
\end{eqnarray}
where
\begin{widetext}
\begin{eqnarray}
{\cal E}= \frac{1}{2 A^2}\left[\frac{\ddot R}{R} - \frac{\ddot B}{B} - \left(\frac{\dot R}{R} - \frac{\dot B}{B}\right)\left(\frac{\dot A}{A} + \frac{\dot R}{R}\right)\right]+ \frac{1}{2 B^2} \left[\frac{A^{\prime\prime}}{A} - \frac{R^{\prime\prime}}{R} + \left(\frac{B^{\prime}}{B} + \frac{R^{\prime}}{R}\right)\left(\frac{R^{\prime}}{R}-\frac{A^{\prime}}{A}\right)\right] - \frac{1}{2 R^2}.
\label{E}
\end{eqnarray}
\end{widetext}

Observe that we may also write $E_{\alpha\beta}$ as:
\begin{equation}
E_{\alpha \beta}={\cal E} (\chi_\alpha
\chi_\beta-\frac{1}{3}h_{\alpha \beta}). \label{52}
\end{equation}

Finally, using the field equations the following expression may be obtained for  ${\cal E}$ (see \cite{she} or \cite{inh} for details),
\begin{equation}
{\cal E}= -4\pi \Pi+\frac{4\pi}{R^3}\int^r_0{R^3 \mu^\prime d\tilde r} -\frac{12\pi}{R^3}\int^r_0{qUBR^2 d\tilde r}. \label{E}
\end{equation}

\section{An evolution  equation for ${\cal E}$}

As mentioned in the Introduction a differential equation for the Weyl tensor  plays a central role in our work;  this equation which follows from the Bianchi identities, was  originally found in \cite{ellis1, ellis2} and was reobtained  in \cite{12}.  Here we  use the notation used in \cite{inh}; it reads:

\begin{widetext}
\begin{eqnarray}
\frac{\partial}{\partial t}\left[{\cal E }- 4\pi\left( \mu - \Pi \right)\right]
= \frac{3\dot R}{R}\left[4 \pi \left( \mu + P_\bot \right)-{\cal E}\right] + 12 \pi  q \frac{A R^\prime}{B R}.
\label{6}
\end{eqnarray}
\end{widetext}

In the next section we  elaborate on this equation, rewriting it in such a way that it may be regarded as an evolution equation for the anisotropy $\Pi$,  thereby  providing the conditions  ensuring the  propagation in time of pressure isotropy. 

\section{The evolution of the pressure isotropy condition}
Let us start our discussion by noticing a fact which is seldom mentioned in the study of  relativistic hydrodynamics; we have in mind the ``asymmetry''  in the role played by   the radial and tangential pressure in the context of general relativity. Indeed, in the static case the TOV equation may be written at once from (\ref{j5}) as
\begin{eqnarray}
 P_r^{\prime}
+\left(\mu+ P_r \right)\frac{A^{\prime}}{A}
+2( P_r-P_{\perp})\frac{R^{\prime}}{R}=0. \label{j5b}
\end{eqnarray}

The above equation is the hydrostatic equilibrium equation and the physical meaning of its different terms is well known: the first term is just the gradient of pressure opposing gravity, the second term describes the gravitational ``force''  and finally the third term describes the effect of the pressure anisotropy, whose sign  depends on the difference between principal stresses. 

The remarkable fact is that  while the radial pressure enters into the gravitational force term, the tangential pressure does not. In other words there is not a ``self--regenerative effect '' of the tangential pressure, which explains why anisotropic spheres may be more compact than isotropic ones (if $P_{\perp}>P_r$). This is a purely relativistic effect, since in the Newtonian limit the radial pressure  in the second term of (\ref{j5b}) vanishes and both principal stresses appear symmetrically in the hydrostatic equilibrium equation. In other words, in relativistic hydrodynamics there seems to be an {\it intrinsic} anisotropy, in the sense that the role played by principal stresses is different.

A hint about the origin of the departure from the pressure isotropy condition during the evolution is provided by the following ``qualitative'' analysis of  a system leaving the equilibrium from a static fluid distribution with isotropic pressure.

Thus, let us assume that our system is forced to abandon the state of equilibrium, and we take a ``snapshot'' of the system immediately after that, at a time scale shorter than the thermal relaxation time, the thermal  adjustment time  and the hydrostatic time. Therefore, at this time scale we have:
\begin{equation}
q \approx U \approx \Theta \approx  \sigma \approx 0\Rightarrow \dot R \approx  \dot B \approx 0,
\label{l1}
\end{equation}
obviously the time derivatives of the above quantities are small but nonvanishing.

Then, evaluating the anisotropic scalar $\Pi$ at this time scale, we obtain from (\ref{14}) and (\ref{15})
\begin{equation}
8\pi \Pi\approx \frac{1}{A}\left(\frac{\ddot B}{B}-\frac{\ddot R}{R}\right)\approx \dot \sigma,
\label{l2}
\end{equation}

where the fact has been assumed that the fluid is initially isotropic in the pressure.

Thus it appears that unless we assume that the fluid evolves shear--free, at least within the time scale under consideration, it will depart from the initial isotropic pressure condition. It might be argued that for some unknown physical reasons, some ``isotropization'' process brings the system back to the isotropic pressure condition. However  this is a highly speculative assumption and the fact remains that the expected tendency of the system is to develop pressure anisotropy.

This result, although valid only for the time scale under consideration, should not be underestimated.
Indeed, once the system is removed from equilibrium, it faces two possible scenarios: a) the fluid is stable and   gets  back to a static  regime  within a time scale of the order of hydrostatic time, or,  b) it is unstable, and enters into a dynamic regime until eventually  reaching  a  final equilibrium state. 
In the former case a), there is no reason to think that the acquired anisotropy given by  (\ref{l2}) would disappear in the new equilibrium state, and therefore the resulting configuration, unlike the initial one, even if it is static  should in principle exhibit,  pressure anisotropy.

In the latter case b), we see next  that the departure from the isotropic pressure condition is the rule, for any time scale, even if we assume that the evolution proceeds shear--free.

For doing so we shall elaborate on (\ref{6}) as follows. 

Using (\ref{j4}) and (\ref{5b1}) we may write (\ref{6}) in the form,
\begin{widetext}
\begin{eqnarray}
\frac{\partial}{\partial t}\left({\cal E }+4\pi\ \Pi \right)+ \frac{\dot R}{R}\left(3 {\cal E }+4\pi \Pi \right) =-4 \pi \left( \mu + P_r \right)A \sigma  -\frac{4\pi q}{B}\left(2A^\prime -\frac{A R^\prime}{ R}\right)-\frac{4\pi q^\prime A}{B},
\label{6l}
\end{eqnarray}
\end{widetext}
or introducing for simplicity the dissipative factor ($\Psi_{diss.}$),
\begin{eqnarray}
\Psi_{diss} \equiv -\frac{4\pi}{B}\left[\left(2A^\prime -\frac{A R^\prime}{ R}\right)q+q^\prime A,\right],
\label{7l}
\end{eqnarray}
we may rewrite (\ref{6l}) as an evolution equation for the anisotropy $\Pi$, as 
\begin{widetext}
\begin{eqnarray}
 \dot \Pi + \frac{\dot R}{R}\Pi+\frac{1}{4\pi}\left( \dot {\cal E }+\frac{3{\cal E}\dot R}{R}\right) =-\left( \mu + P_r\right)A \sigma  +\frac{1}{4\pi}\Psi_{diss.}.
\label{8l}
\end{eqnarray}
\end{widetext}

The  above equation may be integrated, producing,
\begin{widetext}
\begin{eqnarray}
 \Pi =-\frac{1}{4\pi R} \int^t_0{R \left( \dot {\cal E }+\frac{3{\cal E}\dot R}{R}\right)d\tilde t} -\frac{1}{R}\int^t_0{\left( \mu + P_r \right)A \sigma Rd\tilde t} +\frac{1}{4\pi R}\int^t_0{R\Psi_{diss.}d\tilde t},
\label{9l}
\end{eqnarray}
\end{widetext}
where the initial condition $\Pi(t=0)=0$ has been imposed.

At this stage, we may identify in the equation above three different factors forcing the system to abandon the pressure isotropy condition. The first integral provides the contribution from the Weyl tensor, the second one depends on the shear of the flow and the last one describes the role played by the dissipative processes through the dissipative factor.

We next transform the equation above by expressing the Weyl tensor terms in the first integral, through its expression (\ref{E}).

Thus using (\ref{E})  in (\ref{9l}), we obtain after some simple calculations
\begin{eqnarray}
\Pi \dot R&=&\frac{R}{2}\left(\mu+P_r\right) A\sigma -\frac{R \Psi_{diss.}}{8\pi}-\frac{3}{2R^2}\frac{\partial}{\partial t}\left(\int^r_0{qUBR^2d\tilde r}\right)\nonumber \\&+&\frac{1}{2R^2}\frac{\partial}{\partial t}\left(\int^r_0{R^3\mu^\prime d\tilde r}\right).
\label{10l}
\end{eqnarray}

We see from the above equation that unless a highly unlikely cancellation of the four terms on the right occurs, the system will abandon the pressure isotropic condition.

We shall next analyze the axially symmetric dissipative case.

\section{The axially symmetric case }
A general approach to analyze axially  and reflection symmetric fluids was  developed in \cite{1} based on the $1+3$ formalism \cite{ellis1,ellis2,ellis3}. Thus it is not  difficult to realize that the analysis presented in the previous section could be extended to the axially symmetric case, by using  equations (B10)--(B13) in \cite{1}. However such analysis would involve extremely long expressions making it difficult  to extract a useful information.  Instead, still using the results of \cite{1}, we present in this section  a more qualitative  approach, which however   provides enough arguments as to consider the departure from the pressure isotropy as the rule instead of the exception, in this case too. 

More specifically, as we already did at the beginning of the previous section,  we  analyze the  behaviour of the system immediately after its  departure from equilibrium.
By   ``immediately'' we mean at the smallest time scale at which we can observe the first signs of dynamical evolution. Such a time scale is assumed to be smaller than the thermal relaxation time, the hydrostatic time, and the thermal adjustment time.

Thus, we consider,  axially (and reflection) symmetric sources. For such  systems the  line element may be written  as:

\begin{equation}
ds^2=-A^2 dt^2 + B^2 \left(dr^2
+r^2d\theta^2\right)+C^2d\phi^2+2Gd\theta dt, \label{1b}
\end{equation}
where $A, B, C, G$ are positive functions of $t$, $r$ and $\theta$. We number the coordinates $x^0=t, x^1=r, x^2= \theta, x^3=\phi$.

We  assume that  our source is filled with an anisotropic and dissipative fluid. 

The energy momentum tensor may be written in the canonical  form, as 
\begin{equation}
{T}_{\alpha\beta}= (\mu+P) V_\alpha V_\beta+P g _{\alpha \beta} +\Pi_{\alpha \beta}+q_\alpha V_\beta+q_\beta V_\alpha.
\label{6bis}
\end{equation}

Choosing the fluid to be comoving in our coordinates, then
\begin{equation}
V^\alpha =\left(\frac{1}{A},0,0,0\right); \quad  V_\alpha=\left(-A,0,\frac{G}{A},0\right).
\label{m1}
\end{equation}

We  next define a canonical  orthonormal tetrad (say  $e^{(a)}_\alpha$), by adding to the four--velocity vector $e^{(0)}_\alpha=V_\alpha$, three spacelike unitary vectors (these correspond to the vectors $\bold K, \bold L, \bold S$ in \cite{1})

\begin{equation}
e^{(1)}_\alpha=(0,B,0,0); \quad  e^{(2)}_\alpha=\left(0,0,\frac{\sqrt{A^2B^2r^2+G^2}}{A},0\right),
\label{7}
\end{equation}

\begin{equation}
 e^{(3)}_\alpha(0,0,0,C),
\label{3nb}
\end{equation}
with $a=0,\,1,\,2,\,3$ (latin indices labeling different vectors of the tetrad).

Then the anisotropic tensor  may be  expressed through three scalar functions defined as (see \cite{ax2} for details):

\begin{eqnarray}
 \Pi _{(2)(1)}=e^\alpha_{(2)}e^\beta_{(1)} T_{\alpha \beta} 
, \quad  \label{7P}
\end{eqnarray}

\begin{equation}
\Pi_{(1)(1)}=\frac{1}{3}\left(2e^{\alpha}_{(1)} e^{\beta}_{(1)} -e^{\alpha}_{(2)} e^{\beta}_{(2)}-e^{\alpha}_{(3)} e^{\beta}_{(3)}\right) T_{\alpha \beta},
\label{2n}
\end{equation}
\begin{equation}
\Pi_{(2)(2)}=\frac{1}{3}\left(2e^{\alpha}_{(2)} e^{\beta}_{(2)} -e^{\alpha}_{(3)} e^{\beta}_{(3)}-e^{\alpha}_{(1)} e^{\beta}_{(1)}\right) T_{\alpha \beta}.
\label{2nbis}
\end{equation}

The heat flux vector may be defined in terms of  the two tetrad components $q_{(1)}$ and $q_{(2)}$, as:
\begin{equation}
q_\mu=q_{(1)}e_{\mu}^{(1)}+q_{(2)}e_{\mu}^{(2)}
\label{qn1}
\end{equation}
or, in coordinate components (see \cite{1})
\begin{equation}
q^\mu=\left(\frac{q_{(2)} G}{A \sqrt{A^2B^2r^2+G^2}},  \frac{q_{(1)}}{B}, \frac{Aq_{(2)}}{\sqrt{A^2B^2r^2+G^2}}, 0\right)
,\label{q}
\end{equation}
\begin{equation}
 q_\mu=\left(0, B q_{(1)}, \frac{\sqrt{A^2B^2r^2+G^2}q_{(2)}}{A}, 0\right).
\label{qn}
\end{equation}

The four acceleration, may be expressed through two scalar functions 
\begin{equation}
a_\alpha=V^\beta V_{\alpha;\beta}=a_{(1)}e_{\mu}^{(1)}+a_{(2)}e_{\mu}^{(2)},
\label{a1n}
\end{equation}
with
\begin{equation}
a_{(1)}= \frac {A^\prime }{AB };\quad a_{(2)}=\frac{A}{\sqrt{A^2B^2r^2+G^2}}\left[\frac {A_{,\theta}}{A}+\frac {G}{A^2}\left(\frac{\dot G}{G}-\frac{\dot A}{A}\right)\right],
\label{acc}
\end{equation}
 where the dot  and the prime denote derivatives with respect to $t$ and $r$ respectively. 

For the expansion scalar we obtain 
\begin{eqnarray}
\Theta&=&V^\alpha_{;\alpha}=\frac{1}{A}\left(\frac{2 \dot B}{B}+\frac{\dot C}{C}\right) \nonumber\\
&+&\frac{G^2}{A\left(A^2 B^2 r^2 + G^2\right)}\left(-\frac{\dot A}{A}-\frac{\dot B}{B}+\frac{\dot G}{G}\right),
\label{theta}
\end{eqnarray}

whereas, the shear tensor is defined in terms of two scalar functions $\sigma_{(1)(1)}$ and $\sigma_{(2)(2)}$, 
which may be written in terms of the metric functions and their derivatives as (see \cite{1}):
\begin{eqnarray}
\sigma_{(1) (1)}&=&\frac{1}{3A}\left(\frac{\dot B}{B}-\frac{\dot C}{C}\right)\nonumber \\
&+&\frac{G^2}{3A\left(A^2 B^2 r^2 + G^2\right)}\left(\frac{\dot A}{A}+\frac{\dot B}{B}-\frac{\dot G}{G}\right),
 \label{sigmasI}
\end{eqnarray}
\begin{eqnarray}
\sigma_{(2)(2)}&=&\frac{1}{3A}\left(\frac{\dot B}{B}-\frac{\dot C}{C}\right)\nonumber \\ &+&\frac{2G^2}{3A\left (A^2 B^2 r^2 + G^2\right)}\left(-\frac{\dot A}{A}-\frac{\dot B}{B}+\frac{\dot G}{G}\right)
\label{sigmas}.
\end{eqnarray}

Finally,  for  the vorticity tensor 
\begin{equation}
\Omega
_{\beta\mu}=\Omega_{(a)(b)}e^{(a)}_\beta
e^{(b)}_\mu,
\end{equation}
we find that it is determined by a single basis component:
\begin{equation}
\Omega_{(1)(2)} = -\Omega_{(2)(1)}=-\Omega=-\frac{G(\frac{G^\prime}{G}-\frac{2A^\prime}{A})}{2B\sqrt{A^2B^2r^2+G^2}}.
\label{omegan}
\end{equation}
 
It is important to recall  that we have to impose regularity conditions, necessary to ensure elementary flatness in the vicinity of  the axis of symmetry, and in particular at the center (see \cite{1n}, \cite{2n}, \cite{3n}), thus   as $r\approx 0$
\begin{equation}
\Omega=\sum_{n \geq1}\Omega^{(n)}(t,\theta) r^{n},
\label{sum1}
\end{equation}
implying, because of (\ref{omegan}) that in the neighborhood of the center
\begin{equation}
G=\sum_{n\geq 3}G^{(n)}(t, \theta) r^{n}.
\label{sum1}
\end{equation}
Next, we  need a transport equation, here we use the  M\"{u}ller-Israel-Stewart second
order phenomenological theory for dissipative fluids \cite{18, 19, 20, 21}). However, the main conclusions generated by our analysis  are not dependent on the transport equation  chosen, as far as it is a causal one, i.e that it leads to a  Cattaneo type  equation \cite{70}, leading
thereby to a hyperbolic equation for the propagation of thermal
perturbations.

Thus, the transport equation for the heat flux reads \cite{19, 20, 21},
\begin{equation}
\tau h^\mu_\nu q^\nu _{;\beta}V^\beta +q^\mu=-\kappa
h^{\mu\nu}(T_{,\nu}+T a_\nu)-\frac{1}{2}\kappa T^2\left
(\frac{\tau V^\alpha}{\kappa T^2}\right )_{;\alpha}q^\mu,\label{qT}
\end{equation}

\noindent where $\tau$, $\kappa$, $T$ denote the relaxation time,
the thermal conductivity and the temperature, respectively.
Contracting (\ref{qT}) with $e^{(2)}_\mu$ we obtain
\begin{widetext}
\begin{eqnarray}
\frac{\tau}{A}\left(\dot q_{(2)}+A q_{(1)} \Omega\right)+q_{(2)}=-\frac{\kappa}{A}\left(\frac{G \dot T+A^2 T_{,\theta}}{\sqrt{A^2B^2r^2+G^2}}+A T a_{(2)}\right) -\frac{\kappa T^2q_{(2)}}{2}\left(\frac{\tau V^\alpha}{\kappa T^2}\right)_{;\alpha},\label{qT1n}
\end{eqnarray}
\end{widetext}
whereas   the contraction of  (\ref{qT}) with  $e^{(1)}_\mu$, produces

\begin{eqnarray}
\frac{\tau}{A}\left(\dot q_{(1)}-A q_{(2)} \Omega\right)+q_{(1)}=-\frac{\kappa}{B}\left(T^\prime+BTa_{(1)}\right)\nonumber \\
-\frac{\kappa T^2 q_{(1)}}{2}\left(\frac{\tau
V^\alpha}{\kappa T^2}\right)_{;\alpha}. \label{qT2n}
\end{eqnarray}

Let us now take a snapshot of the system, just after it has abandoned the equilibrium. As mentioned before, by ``just after'' we mean on the smallest time scale, at which we can detect the first signs of dynamical evolution. 
The following results follow from this evaluation (see \cite{ax2} for details).
\begin{itemize}
\item At the time scale at which we are observing the system, which is smaller than the hydrostatic time scale,  the kinematical quantities $\Omega (G), \Theta, \sigma_{(1) (1)},\sigma_{(2) (2)}$  keep the same values they have in equilibrium, i.e. they are neglected (of course not so their time derivatives which are assumed to be small, say of order $O(\epsilon)$, (where $\epsilon<<1$), but non--vanishing.
\item The heat flux vector should also be   neglected (once again, not so its time derivative). 
\item From the above conditions it follows at once that  first order time derivatives of the metric variables $A, B, C$ can be neglected.
\end{itemize}

Then, we have for the four acceleration
\begin{equation}
a_{(1)}= \frac {A^\prime }{AB };\quad a_{(2)}=\frac{1}{Br}\left(\frac {A_{,\theta}}{A}+\frac{\dot G}{A^2}\right),
\label{accitem}
\end{equation}

and from the remaining kinematical variables
\begin{equation}
\dot \Theta=\frac{1}{A}\left(\frac{2\ddot B}{B}+\frac{\ddot C}{C}\right),\quad\dot \sigma_{(1) (1)}=\dot \sigma_{(2) (2)}\equiv  \dot {\bar \sigma}=\frac{1}{3A}\left(\frac{\ddot B}{B}-\frac{\ddot C}{C}\right),
\label{ds}
\end{equation}
\begin{equation}
\dot \Omega=\frac{1}{AB^2r}\left(\frac{\dot G^{\prime}}{2}-\frac{\dot G A^{\prime}}{A}\right).
\label{vornn}
\end{equation}

Now, at thermal equilibrium,  when the heat flux vanishes, the Tolman conditions for thermal equilibrium  \cite{Tolman}
\begin{equation}
(TA)^\prime=(TA)_{,\theta}=0,
\label{tol}
\end{equation}
are valid. 

Thus, the evaluation of (\ref{qT2n})  and (\ref{qT1n}) just after leaving the equilibrium, produces respectively

\begin{eqnarray}
\dot q_{(1)}=0 ,\label{qT1nn}
\end{eqnarray}
and

\begin{eqnarray}
\tau \dot q_{(2)}=-\frac{\kappa AT_{,\theta}}{Br}-\kappa ATa_{(2)},
 \label{qT2nnn}
\end{eqnarray}
or, using (\ref{tol})
\begin{eqnarray}
\tau \dot q_{(2)}=-\frac{\kappa T \dot G}{ABr}
. \label{qT2nn}
\end{eqnarray}

Therefore, at the very beginning of  the evolution, the dissipative process starts with contributions along the $e^{(2)}_\mu$ (meridional) direction.

With the information above we may  calculate the components of the Einstein tensor $G_{\alpha \beta}$ and evaluate them just after the system leaves the equilibrium. At this time scale, this tensor has  three types of terms: On the one hand, there are terms with first time derivatives of the metric functions $A, B, C$, which   are set to zero, next, there are terms  that neither contain $G$, nor first time derivatives of  $A, B, C$,  these  correspond to the expressions in equilibrium, and finally, there are terms with first time derivatives of $G$ and/or  second time derivatives of $A, B, C$,  which of course are not neglected. Then it follows from  the  Einstein equations (see \cite{ax2} for details), 

\begin{equation}
8\pi \mu=8\pi\mu_{(eq)},
\label{moeq}
\end{equation}
\begin{equation}
8\pi P=8\pi P_{(eq)}-\frac{2}{3A}\dot \Theta+\frac{2}{3A^2B^2r^2}\left(\dot G_{,\theta}+\dot G\frac{C_{,\theta}}{C}\right),
\label{Poeq}
\end{equation}
\begin{eqnarray}
&&8\pi \Pi_{(1) (1)}=8\pi \Pi_{(1)(1) (eq)}+\frac{\dot {\bar \sigma}}{A}\nonumber \\&+&\frac{1}{3A^2B^2r^2}\left[\dot G_{,\theta}-\dot G\left(\frac{3B_{,\theta}}{B}-\frac{C_{,\theta}}{C}\right)\right],
\label{PiIoeq}
\end{eqnarray}
\begin{eqnarray}
&&8\pi \Pi_{(2)(2)}=8\pi \Pi_{(2)(2) (eq)}+\frac{\dot {\bar \sigma}}{A}\nonumber \\&+&\frac{1}{3A^2B^2r^2}\left[-2\dot G_{,\theta}+\dot G\left(\frac{3B_{,\theta}}{B}+\frac{C_{,\theta}}{C}\right)\right],
\label{PiIIoeq}
\end{eqnarray}
\begin{equation}
8\pi \Pi_{(2)(1)}=8\pi \Pi_{(2)(1)(eq)}-\frac{\dot \Omega}{A}+\frac{\dot G}{A^2B^2r}\left[\frac{(Br)^\prime}{Br}-\frac{A^\prime}{A}\right],
\label{PiKLoeq}
\end{equation}
where $(eq)$ stands for the value of the quantity at equilibrium.

Now, let us assume that initially the pressure of the system is isotropic  i.e. $\Pi_{(1)(1) (eq)}=\Pi_{(2)(2)(eq)}$ and $\Pi_{(2)(1)(eq)}=0$. The fundamental question  we have to answer  is: may these conditions propagate in time?  To simplify the discussion, let us  assume that  out of equilibrium, at the time scale considered here, we still have $\Pi_{(1)(1)} =\Pi_{(2)(2)}$, then it follows from ( \ref{Poeq}, \ref{PiIoeq}),  that:
\begin{equation}
\dot G=B^2f(t,r),
\label{bnf}
\end{equation}
which by an appropriate choice of the arbitrary function $f$ (referred to as the fluid news function in \cite{ax2}), satisfies regularity conditions and is not in contradiction with any of the equations describing the system.
Thus in principle we may assume that once the system abandons the equilibrium, it may keep (at the time scale under consideration) the condition $\Pi_{(1)(1)} =\Pi_{(2)(2)}$.
However, the situation is quite different for  the scalar $\Pi_{(2)(1)}$. In fact,  if we impose the condition  $\Pi_{(2)(1)}=0$, then 
 because of (\ref{PiKLoeq}), we have
\begin{equation}
\frac{\dot \Omega}{A}=\frac{\dot G}{A^2B^2r}\left[\frac{(Br)^\prime}{Br}-\frac{A^\prime}{A}\right],
\label{PiKLoeqb}
\end{equation}
which together with (\ref{omegan}) produces
\begin{equation}
\dot G=B^2r^2g(t,\theta),
\label{bnfbis}
\end{equation}
where $g$ is an arbitrary function of its arguments. But, (\ref{bnfbis}) clearly violates the regularity condition  (\ref{sum1}), close to the center. Accordingly, at the time scale under consideration we must have $\Pi_{(2)(1)}\neq 0$, more precisely
\begin{equation}
8\pi \Pi_{(2)(1)}=\frac{f(t,r)}{2A^2r}\left(\ln{\frac{r^2}{f}}\right)^\prime.
\label{PiKLoeqbb}
\end{equation}
Thus we see that, after leaving the equilibrium, at the time scale under consideration,  the condition  $\Pi_{(1)(1)} =\Pi_{(2)(2)}$ may be assumed to hold. However for the off diagonal  tension $\Pi_{(2)(1)}$ the situation is quite different, at our time scale. Indeed, the function $f$ controls the evolution of the system as it abandons the equilibrium, accordingly it must be different from zero, and so should   $\Pi_{(2)(1)}$, even if we assume it to vanish initially. 

It is worth emphasizing that a non--vanishing  function $f$,  triggers the onset of dissipative processes as it follows from (\ref{qT2nn}), and the appearance of shear, according to Equation (62) in  \cite{ax2}. Additionally, as shown in \cite{1} (section X), the dissipative processes are responsible (among other factors) for  the appearance of energy--density inhomogeneities. Therefore, as in the spherically symmetric case, here too, the above mentioned physical factors   bring on  the onset of pressure anisotropy as the system exits from the  initial state with isotropic pressure.

Although the above argument is valid only for the time scale under consideration, it nevertheless  brings out the tendency of the system to develop pressure anisotropy during the evolution. More so, it should be stressed, as we did in the spherically symmetric case, that if the system returns to a static regime within a time scale of the order of hydrostatic time scale, it will do so keeping the non vanishing value of $\Pi_{(2)(1)}$  acquired after leaving the equilibrium, i.e. in the new static regime the fluid would be anisotropic.

\section{Discussion}
Fundamental results have been obtained during the last decades, concerning  Newtonian and relativistic fluids,  under the assumption that the pressure is isotropic. However we know that the appearance of small amounts of pressure anisotropy  may be enough to produce quite different results, under otherwise the same general conditions. Also, we know that many physical processes producing pressure anisotropy are expected to be present in very compact objects. Based on these comments we felt compelled to pose the following question: under which conditions would an initial fluid configuration with isotropic pressure remain so during its evolution?

For the spherically symmetric case the qualitative analysis presented at the beginning of section IV, points out the tendency of the system to abandon the isotropic pressure condition (at least for a specific time scale). Furthermore,  this result strongly suggests that if the system is stable and gets back to equilibrium after having been removed from it, in this new state of equilibrium the fluid would be anisotropic. The more rigorous analysis presented next, confirmed this tendency for an arbitrary scale time, and allows us to identify the physical factors  inducing the appearance of pressure anisotropy. According to (\ref{10l}) these factors are: the shear, the heat flux vector through the dissipative factor and the first integral in (\ref{10l}), and  the energy  density inhomogeneity. This point deserves a deeper analysis. 

Indeed, as  is apparent from (\ref{10l}), in order for  an initial fluid configuration with isotropic pressure to remain isotropic all along the evolution, we must require  the fluid to be nondissipative, shear--free and homogeneous in the energy--density, unless we admit the highly  unlikely cancelation of the four terms  on the right of (\ref{10l}).  Alternatively, as can be seen form (\ref{9l}), the conditions ensuring the stability of the pressure isotropic condition are, conformal flatness, vanishing shear and absence of dissipation, unless,  again, we assume  the exceptional cancellation of the three terms on the right of  (\ref{9l}).

Now, it has been  shown in \cite{inh} that  the shear,  and/or local anisotropy of pressure and/or dissipative fluxes entail the formation of energy density inhomogeneities.  On the other hand it has been shown in \cite{she} that  the departure from the shear--free condition is controlled by  a single scalar function defined in terms of  the anisotropy of the pressure, the dissipative variables and the energy density inhomogeneity. Thus even if we assume that initially not only the pressure anisotropy but also the shear and the energy--density inhomogeneity vanishes, the dissipative flux would enhance the departure of the isotropic pressure condition through two different channels: on the one hand by its contribution as described by  (\ref{10l}), and on the other hand by inducing departures from the shear--free condition and energy density homogeneity. Thus only imposing  the conformal flatness, the nondissipation, and the shear--free  conditions all along the evolution, can we ensure that the fluid  evolves keeping  the isotropic pressure condition at all times. However, it is worth recalling that   dissipation due to the emission of massless particles (photons and/or neutrinos)  is the only plausible mechanism to carry away the bulk of the huge binding energy of the collapsing star, leading to a neutron star or black hole. In other words, the adiabatic condition imposed to a collapsing scenario  is very unrealistic and dissipation has to be taken  into account in any physically meaningful description of stellar evolution, thereby entailing  the departure from the isotropic pressure condition.

In section VI we analyzed the same question for axially symmetric fluid systems. However, for simplicity, we did not deduce the explicit equation of evolution for the anisotropic scalars, but,  instead, considered the system at the shortest time scale at which the first signs of dynamical evolution can be observed. It was shown that starting from an initially isotropic fluid, at the time scale under consideration, the evolution leads to an anisotropic fluid.  Indeed, it appears  that even if we assume that the two anisotropic scalar functions $\Pi_{(1)(1)}$  and $\Pi_{(2)(2)}$ remain equal after the departure from equilibrium,  the third anisotropic scalar  $\Pi_{(2)(1)}$ must be necessarily  different from zero after leaving the initial state. Also, as in the spherically symmetric case,  shear, dissipative processes and energy--density inhomogeneities are related to the  onset of pressure anisotropy. Once again, since we expect that the final stages of stellar evolution should be accompanied by intense dissipative processes, we should expect some degree of pressure anisotropy to appear in the nonspherical collapse too.

To summarize: what we have learned so far is that an initial fluid configuration with isotropic pressure would  tend to develop pressure anisotropy as it evolves, under conditions expected in stellar evolution. Of course the magnitude of the acquired pressure anisotropy would depend on the specific data of the system.  However the obtained result allows us to conclude that as well as it is wise to check the stability of any specific result obtained under the assumption of the isotropic pressure condition against fluctuations of this latter condition, it  would also be wise to check the stability of the condition itself, in each case.

\begin{acknowledgments}
This work was partially
supported by Ministerio de Ciencia, Innovacion y Universidades. Grant number: PGC2018--096038--B--I00.
\end{acknowledgments}
\appendix
\section{Einstein equations and conservation equations for the spherically symmetric case}
 Einstein's field equations for the interior spacetime (\ref{1m}) are given by
\begin{equation}
G_{\alpha\beta}=8\pi T_{\alpha\beta},
\label{2}
\end{equation}
and its non zero components
with (\ref{1m}) and (\ref{3})
become
\begin{widetext}
\begin{eqnarray}
8\pi T_{00}=8\pi  \mu A^2
=\left(2\frac{\dot{B}}{B}+\frac{\dot{R}}{R}\right)\frac{\dot{R}}{R}
-\left(\frac{A}{B}\right)^2\left[2\frac{R^{\prime\prime}}{R}+\left(\frac{R^{\prime}}{R}\right)^2
-2\frac{B^{\prime}}{B}\frac{R^{\prime}}{R}-\left(\frac{B}{R}\right)^2\right],
\label{12} \\
8\pi T_{01}=-8\pi qAB
=-2\left(\frac{{\dot R}^{\prime}}{R}
-\frac{\dot B}{B}\frac{R^{\prime}}{R}-\frac{\dot
R}{R}\frac{A^{\prime}}{A}\right),
\label{13} \\
8\pi T_{11}=8\pi P_r B^2 
=-\left(\frac{B}{A}\right)^2\left[2\frac{\ddot{R}}{R}-\left(2\frac{\dot A}{A}-\frac{\dot{R}}{R}\right)
\frac{\dot R}{R}\right]
+\left(2\frac{A^{\prime}}{A}+\frac{R^{\prime}}{R}\right)\frac{R^{\prime}}{R}-\left(\frac{B}{R}\right)^2,
\label{14} \\
8\pi T_{22}=\frac{8\pi}{\sin^2\theta}T_{33}=8\pi P_{\perp}R^2
=-\left(\frac{R}{A}\right)^2\left[\frac{\ddot{B}}{B}+\frac{\ddot{R}}{R}
-\frac{\dot{A}}{A}\left(\frac{\dot{B}}{B}+\frac{\dot{R}}{R}\right)
+\frac{\dot{B}}{B}\frac{\dot{R}}{R}\right]\nonumber \\
+\left(\frac{R}{B}\right)^2\left[\frac{A^{\prime\prime}}{A}
+\frac{R^{\prime\prime}}{R}-\frac{A^{\prime}}{A}\frac{B^{\prime}}{B}
+\left(\frac{A^{\prime}}{A}-\frac{B^{\prime}}{B}\right)\frac{R^{\prime}}{R}\right].\label{15}
\end{eqnarray}
\end{widetext}

The nontrivial components of the Bianchi identities, $T^{\alpha\beta}_{;\beta}=0$, from (\ref{2}) yield
\begin{widetext}
\begin{eqnarray}
T^{\alpha\beta}_{;\beta}V_{\alpha}=-\frac{1}{A}\left[\dot { \mu}+
\left( \mu+ P_r\right)\frac{\dot B}{B}
+2\left( \mu+P_{\perp}\right)\frac{\dot R}{R}\right] 
-\frac{1}{B}\left[ q^{\prime}+2 q\frac{(AR)^{\prime}}{AR}\right]=0, \label{j4}\\
T^{\alpha\beta}_{;\beta}\chi_{\alpha}=\frac{1}{A}\left[\dot { q}
+2 q\left(\frac{\dot B}{B}+\frac{\dot R}{R}\right)\right] 
+\frac{1}{B}\left[ P_r^{\prime}
+\left(\mu+ P_r \right)\frac{A^{\prime}}{A}
+2( P_r-P_{\perp})\frac{R^{\prime}}{R}\right]=0, \label{j5}
\end{eqnarray}
\end{widetext}

\end{document}